\documentstyle[aps,multicol,tabularx,rotate,epsf]{revtex}
\begin{document}
\title{Phonon--Mediated Anomalous Dynamics of Defects}
\author{Ali Najafi and Ramin Golestanian}
\address{Institute for Advanced Studies in Basic Sciences,
Zanjan 45195-159, Iran}
\date{\today}
\maketitle
\begin{abstract}
Dynamics of an array of line defects interacting with a background
elastic medium is studied in the linear regime. It is shown that
the {\em inertial} coupling between the defects and the ambient
phonons leads to an anomalous response behavior for the
deformation modes of a defect-lattice, in the form of {\em
anisotropic} and {\em anomalous} mass and elastic constants,
resonant dissipation through excitation of phonons, and
instabilities. The case of a single fluctuating line defect is
also studied, and it is shown that it could lead to formation of
shock waves in the elastic medium for sufficiently high frequency
deformation modes.

\end{abstract}
\pacs{63.20.Mt,61.72.Bb,66.30.Lw}
\begin{multicols}{2}

Topological defects have a wide range of interest in many branches
of physics. Vortices in superconductors, superfluids, and
Bose--Einstein condensates, crystal defects in solid bodies,
topological excitations in spin systems with continuous symmetry,
and defects in liquid crystals are famous examples of their
ubiquity \cite{Chaikin}. While they play a pivotal role in our
fundamental understanding of the superfluid \cite{KT} and melting
\cite{HNKT} transitions in two dimensions, they are also important
in determining the properties of real materials---for example they
are responsible to a large degree for the plastic behavior of
solid bodies \cite{Ashcroft}.

Over the past decade, the problem of defect dynamics has attracted
a considerable attention mainly in describing the dynamics of
defect--mediated phase transitions \cite{Domb,Blatter,Minn}, and
in determining the physical properties of vortex-lattices in
type-II supercondoctors \cite{Blatter,Nori,Kardar}. However, the
interaction of defects with the excitations of the elastic
background, or phonons, is not taken into account in these
studies, because in the context of equilibrium statistical
thermodynamics the defects are completely separated from the
phonons and there is no net coupling between the two degrees of
freedom \cite{xy}. Quantized phonons, however, are shown to
interact with defects leading to interesting effects
\cite{Sethna2,Sethna,Thouless}. The interesting observation of a
collection of quantized vortices in Bose--Einstein condensates has
also raised new questions about the dynamical behavior of defects
\cite{BE}.

Here, we consider the classical dynamics of a collection of line
defects in the form of a regular array, which are dynamically
fluctuating in an ambient elastic medium away from thermodynamic
equilibrium. The inertial coupling between the defects and the
phonons of the background medium is shown to lead to a plethora of
novel effects in the long time- and length-scale collective
dynamics of the defect-lattice: (i) The deformation modes of the
defect-lattice acquire an anomalous transverse mass, which
diverges for modes that are in resonance with the background
phonons, and a finite longitudinal mass. (ii) There is a
corresponding anomalous transverse elastic constant in the
direction parallel to the line defects, as well as a corresponding
finite longitudinal elastic constant. (iii) An effective anomalous
friction appears at the resonance, signalling the transfer of
mechanical energy from the defect-lattice to the elastic phonons.
(iv) The defect-lattice becomes intrinsically unstable for
frequencies higher than the phononic resonance frequency for each
wavevector, due to the enhanced excitation of phonons. (v) The
elastic moduli in the perpendicular direction are anomalous, and
show an intrinsic instability for the shear modes. The case of a
single fluctuating line defect is also studied. It is shown that
deformation modes along the line defect with a phase velocity
higher than the bulk velocity of phonons lead to the creation of
shock waves. The propagation of phonons in the periodic matrix of
the defects is also studied, and it is proposed that such
arrangements may lead to formation of frequency gaps in the
phononic band structure.

We consider a simple scalar elasticity described by the field
$\theta({\bf r},t)$, which could be a component of the
displacement field in crystals, or the phase of the order
parameter for superfluids. The field is then conveniently
decomposed into two parts as $\theta({\bf r},t)=\theta_{{\rm
ph}}({\bf r},t)+ \theta_{{\rm def}}({\bf r},t)$: (i) a singular
part $\theta_{{\rm def}}({\bf r},t)$ that is a solution of the
defect condition of the form $\oint d{\bf l} \cdot
\nabla\theta=2\pi n$, where the integral is taken over any closed
path around the defect and $n$ is the corresponding winding number
or the topological charge of the defect, and (ii) a nonsingular
part $\theta_{{\rm ph}}({\bf r},t)$ that describes the phononic
degrees of freedom in the elastic medium. While such a
representation assumes a vanishing size for the defects, we should
keep in mind that in reality they always maintain a finite core
size, which is of the order of the atomic lattice constant $a$ in
crystals. With the above decomposition, we then consider an action
as
\begin{equation}
{\cal A}=\int d t d^3{\bf r} \left[\frac{\rho}{2}(\partial_t
{\theta}_{{\rm ph}}+\partial_t{\theta}_{{\rm def}} )^{2}-
\frac{J}{2}(\nabla\theta_{{\rm ph}}+\nabla\theta_{{\rm def}})^2
 \right],\label{action2}
\end{equation}
where the coefficient $\rho$ is a (linear) mass density, and $J$
is a stiffness coefficient. The quantity $c=\sqrt{J/\rho}$ is the
phase velocity of sound waves in the medium.

\begin{figure}
\centerline{ \epsfxsize=7truecm \epsffile{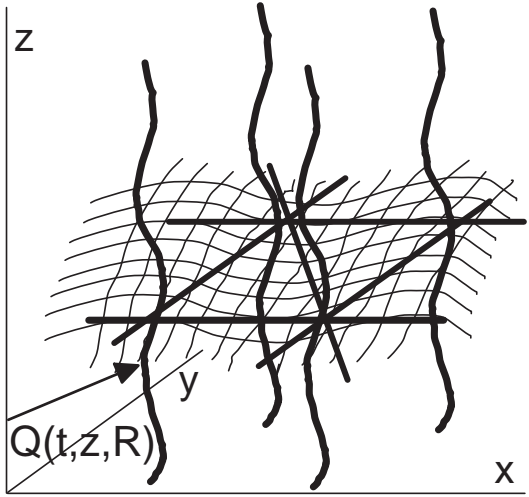} }
\vskip0.3truecm \caption{Array of fluctuating line defects in an
elastic medium. The deformation modes of the defect lattice are
coupled to the phonons of the background medium.} \label{deflat}
\end{figure}

To study the many-body effects in the dynamics of line defects
mediated by phonons, we consider a system consisting of many line
defects (with $n=1$) that are arranged on a triangular lattice
(when at equilibrium) as shown in Fig.~1. In this case the defect
field is given as
\begin{equation}
\theta_{{\rm def}}({\bf r}_{\perp},\{{\bf Q}({\bf
R},z,t)\})=\sum_{\{{\bf R}\}} \arctan \frac {y-Q_2({\bf R},z,t)}
{x-Q_1({\bf R},z,t)}
\end{equation}
where ${\bf r}_{\perp}=(x,y)$, and ${\bf Q}({\bf R},z,t)=(Q_1({\bf
R},z,t),Q_2({\bf R},z,t))$ is the two dimensional position vector
of the core of a line defect in the $x$-$y$ plane, which is
fluctuating around a point $\bf{R}$ of a regular two dimensional
lattice (see Fig.~1). The corresponding Euler-Lagrange equation
for the phonon field then reads
\begin{equation}
\left[\partial^{2}_{t}-c^{2}\nabla^2\right]\theta_{{\rm ph}}({\bf
r},t)=
-\left[\partial^{2}_{t}-c^{2}\partial^{2}_{z}\right]\theta_{{\rm
def}}({\bf r}_{\perp},\{{\bf Q}({\bf R},z,t)\}).\label{ELeq2}
\end{equation}
Note that $\nabla_{\perp}^2 \theta_{{\rm def}}=0$. The above
equation implies that the system of fluctuating defects acts as a
complicated source for phononic waves---each defect can transmit
or absorb phonons from the medium during its motion. We will see
below that the net result of this coupling is a dissipative effect
in which the moving defects lose their energy in the form of
phonon radiation.

To study the collective dynamics of the defects in the presence of
phonons, we should eliminate the phononic degrees of freedom in
the action, and obtain an effective action for the defect degrees
of freedom $\{{\bf Q}({\bf R},z,t)\}$. For this purpose, we assume
that the line defects are only slightly distorted from their
equilibrium configurations, so that their positions can be
described as ${\bf Q}({\bf R},z,t)={\bf R}+{\bf u}({\bf R},z,t)$
with ${\bf u}({\bf R},z,t)$ being much less than the defect
lattice constant $b$. We then solve the wave equation for the
phonons [Eq. (\ref{ELeq2})] perturbatively in the deformation
field ${\bf u}({\bf R},z,t)$, and obtain the effective action up
to the leading order as
\begin{eqnarray}
{\cal A}_{\rm eff}&=&\frac{1}{2}\int_{{\bf q}\in {\rm B.Z.}}
\frac{d^2{\bf q}}{(2\pi)^2} \frac{dk_z}{2\pi}\frac{d\omega}{2\pi}
\;\chi_{\alpha \beta}({\bf q},k_z,\omega) \nonumber \\
&&\times \; u_{\alpha}({\bf q},k_z,\omega) u_{\beta}(-{\bf
q},-k_z,-\omega), \label{Aeff5}
\end{eqnarray}
where
\begin{eqnarray}
\chi_{\alpha \beta}({\bf q},k_z,\omega)&=&J \sum_{{\bf G}}\left\{
\left[\frac{G_\alpha
G_\beta}{G^2}-\frac{(q_\alpha+G_\alpha)(q_\beta+G_\beta)}{({\bf
q}+{\bf G})^2}\right]\right.\nonumber \\
&+&\left.\frac{(\omega^2-c^2 k_z^2)
\left[\frac{(q_\alpha+G_\alpha)(q_\beta+G_\beta)}{({\bf q}+{\bf
G})^2}-\delta_{\alpha\beta}\right]}{\omega^2-c^2 k_z^2-c^2({\bf
q}+{\bf G})^2} \right\}.\label{Kqkw}
\end{eqnarray}
and $\{{\bf G}\}$ represents the reciprocal lattice of the
original defect-lattice $\{{\bf R}\}$. Note that the kernel
consists of two contributions: (i) a term which does not depend on
frequency and comes from the usual long-range logarithmic
interaction between the line defects, and (ii) a term with
dynamical origin describing an additional interaction between the
defects mediated by phonons.

The summation over all the reciprocal lattice vectors of the two
dimensional lattice in the above kernel is difficult to perform
exactly. However, we can study the long time- and length-scale
elasticity of the defect lattice by focusing on the small
frequency and wavevector limit in Eq. (\ref{Kqkw}). This can be
achieved by separating the ${\bf G}=0$ term and expanding the
other terms in powers of $q/G$, $k_z/G$, and $\omega/(c G)$. In
this case the summation over ${\bf G}$ can be performed and we
obtain an analytical expression for the kernel to the leading
order as
\begin{eqnarray}
\chi_{\alpha \beta}({\bf q},k_z,\omega)&=&(M_T
\omega^2-K^{\parallel}_{T} k_z^2-K^{\perp}_{T} q^2)(\delta_{\alpha
\beta}-\hat{q}_{\alpha}\hat{q}_{\beta})\nonumber
\\
&+&(M_L \omega^2-K^{\parallel}_{L} k_z^2-K^{\perp}_{L} q^2)\;
\hat{q}_{\alpha}\hat{q}_{\beta}\nonumber
\\
&+&i \omega \zeta_T \; (\delta_{\alpha
\beta}-\hat{q}_{\alpha}\hat{q}_{\beta}),\label{TL-decom}
\end{eqnarray}
where the effective transverse and longitudinal mass densities are
given as
\begin{equation}
M_T(q,k_z,\omega)={\rho \over q^2+k_z^2-(\omega/c)^2}+{\rho \over
\kappa^2}, \;\;\; M_L={\rho \over \kappa^2},\label{MTL}
\end{equation}
the effective transverse and longitudinal elastic moduli in the
parallel direction as
\begin{equation}
K^{\parallel}_{T}(q,k_z,\omega)={J \over
q^2+k_z^2-(\omega/c)^2}+{J \over
\kappa^2},\;\;\;K^{\parallel}_{L}={J \over
\kappa^2},\label{KparTL}
\end{equation}
the effective transverse and longitudinal elastic moduli in the
perpendicular direction as
\begin{equation}
K^{\perp}_{T}(q)=-{J \over 2 q^2}, \;\;\; K^{\perp}_{L}(q)={J
\over 2 q^2},\label{KperTL}
\end{equation}
and, finally, the effective transverse friction coefficient as
\begin{equation}
\zeta_T(q,k_z,\omega)={\pi J c^2 q^2 \over |\omega|}
\delta(\omega^2-c^2 k_z^2-c^2 q^2).\label{zetaT}
\end{equation}
In the above, parallel and perpendicular are defined with respect
to the line defects, and the {\em screening} length $\kappa^{-1}$
is defined as
\begin{equation}
\kappa^{-2}={1 \over 2}\sum_{{\bf G}\neq 0}\frac{1}{G^2}\simeq
{\pi \ln(b/a)\over G^{*2}},
\end{equation}
where $G^*$ is the lattice constant of the reciprocal lattice,
which for a triangular lattice of line defects is given as
$G^*=\frac{4\pi}{\sqrt{3}b}$. Note that the real (imaginary) part
of the above kernel is even (odd) in $\omega$ and this ensures
that the response is causal.

The above results show that the dynamics of a defect lattice in an
elastic medium is anomalous. While the longitudinal deformation
modes acquire finite mass and elastic modulus in the parallel
direction due to coupling with phonons, the corresponding mass and
elastic modulus in the parallel direction for the transverse modes
are frequency and wavevector dependent, and diverge in the small
wavevector and frequency limit. Moreover, whereas the transverse
mass and elastic modulus in the parallel direction are positive
for $\omega^2 < c^2(q^2+k_z^2)$, they diverge when the frequency
and wavevectors of the deformation modes hit that of a possible
phononic excitation in the background. In this case the
defect-lattice dynamics becomes dissipative---as manifested by the
appearance of an imaginary part in the response kernel that
defines an effective friction coefficient---which corresponds to
transfer of mechanical energy from the defect lattice to the
phonons. For $\omega^2 > c^2(q^2+k_z^2)$, the transverse mass and
elastic modulus in the parallel direction become negative,
signalling an instability in the defect-lattice due to the
resonant coupling with the background phonons. The transverse and
longitudinal elastic moduli in the perpendicular direction are
also wavevector dependent, and they reveal an inherent instability
in the deformation modes that is related to the well-known
instability of a system of charges in electrostatics; often
alluded to as the Earnshaw's theorem \cite{Kardar,Feynman}.

It is also interesting to consider the case where all the line
defects are frozen (to straight lines) and only a single defect is
fluctuating, which can be achieved by assuming that
$u_{\alpha}({\bf q},k_z,\omega)$ does not depend on ${\bf q}$. We
can then combine the restricted ${\bf q}$ integration in the first
Brillouin zone and the summation over the reciprocal lattice
vectors ${\bf G}$ to a free integration over ${\bf q}$, and
calculate the response kernel $\chi_{\rm sl}(k_z,\omega)
\delta_{\alpha \beta}=\int_{\rm B.Z.} \frac{d^2{\bf q}}{(2\pi)^2}
\;\chi_{\alpha \beta}({\bf q},k_z,\omega)$. This yields
\begin{eqnarray}
\chi_{\rm sl}(k_z,\omega)=\rho_{\rm sl} \omega^2-K_{\rm sl}
k_z^2+i \omega \zeta_{\rm sl},\label{chi-sl}
\end{eqnarray}
with the effective linear mass density for a single line defect
given as
\begin{equation}
\rho_{{\rm sl}}(k_z,\omega)=\frac{\pi}{2}\; \rho\; \ln
\left(\frac{\omega_{\rm D}^2}{|\omega^2-c^2
k_z^2|}\right),\label{rho-eff}
\end{equation}
the effective elastic modulus as
\begin{equation}
K_{{\rm sl}}(k_z,\omega)=\frac{\pi}{2}\; J\; \ln
\left(\frac{\omega_{\rm D}^2}{|\omega^2-c^2
k_z^2|}\right),\label{rho-eff}
\end{equation}
and the effective friction coefficient as
\begin{equation}
\zeta_{{\rm sl}}(k_z,\omega)=\frac{\pi^2}{2}\; \rho
\;{(\omega^2-c^2 k_z^2) \over |\omega|} \Theta(\omega^2-c^2
k_z^2).\label{zetakw}
\end{equation}
In the above, the Debye frequency $\omega_{\rm D}\simeq c/a$ is a
high frequency cutoff in the system, and $\Theta$ represents the
Heaviside step function.

The effective mass density and elastic modulus for the line defect
are logarithmically larger than the nominal mass density and
stiffness of the material in the core of the vortex, and they are
also frequency and wavevector dependent. While they are positive
for all frequency and wavevectors, they logarithmically diverge
when $\omega=c k_z$, and in particular in the limit $\omega \to 0$
and $k_z \to 0$.

In the case where the phase velocity of waves on the line defect
$\omega/k_z$ is greater than the phonon velocity in the system
$c$, the response kernel in Eq. (\ref{chi-sl}) develops an
imaginary part, which is a reflection of a transfer of mechanical
energy from the fluctuating defect to the elastic medium in the
form of radiating phonons. This effect is similar to the Cherenkov
radiation, where a charged particle moving in a dielectric medium
with superluminal velocity (i.e. a velocity greater than the
velocity of photons in that medium) radiates electromagnetic
waves.

The time-averaged energy dissipation rate for the fluctuating line
defect can be calculated as
\begin{eqnarray}
P=\int \frac{dk_z}{2\pi}\frac{d\omega}{2\pi} \;\zeta_{{\rm
sl}}(k_z,\omega) \;\omega^2 \; |{\bf u}(k_z,\omega)|^2.
\label{P-1}
\end{eqnarray}
To explicitly check that the dissipation is due to phonon
radiation, we calculate the rate of change in energy for the
phonon field as
\begin{equation}
\frac{dE_{{\rm ph}}}{dt}=\int d^3 {\bf r}\;
\frac{\partial}{\partial t}T_{00},\label{dE/dt-1}
\end{equation}
using the $00$-component of the stress tensor $T_{\mu \nu}$
\cite{Landau2} that can be calculated from the phonon field action
in Eq. (\ref{action2}). The result comes out exactly equal to the
dissipation rate as given in Eq. (\ref{P-1}) above.

To estimate the radiated power, we consider an example in which
the defect is undergoing a solid harmonic motion described via
${\bf u}(k_z,\omega)={\bf u}_0 (2 \pi)^2 \delta(k_z) [\delta
(\omega+\omega_0)+\delta (\omega-\omega_0)]/2$. The time-averaged
radiated power per unit length from this oscillating line defect
is then calculated from Eq. (\ref{P-1}) as $P/L={\pi^2} \rho u_0^2
\omega_0^3/4$.

Finally, we note that the periodic arrangement of the defect lines
on a lattice changes the spectrum of phonons, as in the case of
electrons in metals. This is manifested in the singular points
(poles) in the dynamical term in the response kernel of Eq.
(\ref{Kqkw}) whose structure determines the dispersion relation
for the phononic excitations. The dispersion relation $\Omega({\bf
q})\equiv \sqrt{\omega(q)^2-c^2 k_z^2}=c |{\bf q}-{\bf G}|$ should
respect the symmetry of the reciprocal lattice ${{\bf G}}$. As an
example, the phononic dispersion diagram for the case of a
triangular lattice of line defects is plotted in Fig.~2. The unit
vectors for the triangular lattice are $b(1,0)$ and
$b(\frac{1}{2},\frac{\sqrt{3}}{2})$, and the unit vectors for the
corresponding reciprocal lattice are
$G^*(\frac{\sqrt{3}}{2},-\frac{1}{2})$ and $G^*(0,1)$, where
$G^*=\frac{4\pi}{\sqrt{3}b}$ is the lattice constant of the
reciprocal lattice.

\begin{figure}
\centerline{\epsfxsize=8truecm \epsffile{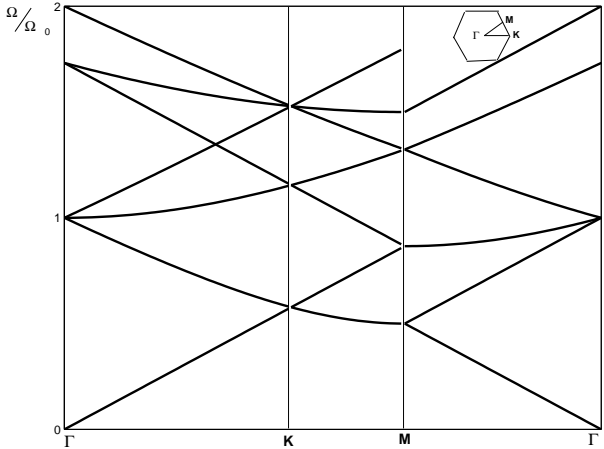}}
\vskip0.3truecm \caption{Diagram of phononic dispersion relation
defined via $\Omega({\bf q})\equiv \sqrt{\omega(q)^2-c^2 k_z^2}=c
|{\bf q}-{\bf G}|$ in a triangular lattice of line defects, with
$\Omega_0=c G^*=\frac{4\pi c}{\sqrt{3}b}$. The inset shows the
first Brillouin zone of the triangular lattice.} \label{band}
\end{figure}

While the band diagram in Fig.~2 has no gaps, one can argue that
an interaction of the phonons with the interior of the defects
that changes the propagation velocity of phonons from $c$ to
$c/\sqrt{\epsilon({\bf r})}$ in the vicinity of the defects
\cite{Minn} can produce a gap in the band diagram, similar to the
case of photonic crystals \cite{Photonic}. A simple calculation
(similar to the estimation of the gap for electrons in a weak
periodic potential \cite{Ashcroft}) then shows that such an
interaction can lead to frequency gaps in the zone boundary of the
order of $\Delta \sim c G^* (a/b)$, where $a$ is the core size of
the defect \cite{Najafi}. This may suggest that a rather dense
array of defects may act as a ``phononic crystal,'' which could be
of potential interest for applications.

In conclusion, we have shown that the dynamical interaction of
defects with phonons can lead to renormalization of mass and
dissipation, which is analogous to the radiation-reaction force
for electrons and their mass renormalization due to their
interaction with electromagnetic fields \cite{Barut}.


We are grateful to J. Sethna for very helpful comments, and to S.
Ramanathan for invaluable discussions during the early stages of
this work.

\end{multicols}
\end{document}